\documentclass[reprint,aps,prd,nofootinbib,floatfix,amsmath,amssymb]{revtex4-2}

\usepackage{amsmath}
\usepackage{amsfonts}
\usepackage{graphicx}
\usepackage{multirow}
\usepackage{cancel}
\usepackage{graphicx}
\usepackage{amssymb}
\usepackage[usenames]{color}
\usepackage[toc,page]{appendix}
\usepackage{slashed}
\usepackage{bbding}
\usepackage{booktabs}

\usepackage[colorlinks=true, urlcolor=blue, linkcolor=blue,
citecolor=blue]{hyperref}

\newcommand{\be}{\begin{equation}}
\newcommand{\ee}{\end{equation}}

\newcommand{\eq}[1]{\begin{align}#1\end{align}}

\newcommand{\dbeta}{$\beta\beta_{0\nu}\,$}
\newcommand{\dlt}{$\Delta L\!=\!2\ $}

\makeatother

\begin{document}

\title{$\Delta$L=2 hyperon decays induced by Majorana neutrinos and doubly-charged scalars}

\author{G. Hern\'andez-Tom\'e}
\affiliation{
Instituto de F\'isica, Universidad Nacional Aut\'onoma de M\'exico, AP 20-364, Ciudad de M\'exico 01000, M\'exico,}
\affiliation{Departamento de F\'isica, Centro de Investigaci\'on y de Estudios Avanzados del Instituto Polit\'ecnico Nacional, Apartado Postal 14-740, 07000 M\'exico D.F., M\'exico}
\author{G. L\'opez Castro}
\author{D.~Portillo-S\'anchez}
\affiliation{Departamento de F\'isica, 
Centro de Investigaci\'on y de Estudios Avanzados del Instituto Polit\'ecnico Nacional,
Apdo.~Postal 14-740, 07000 M\'exico D.F., M\'exico}
\date{\today}

\begin{abstract}
Searches for $\Delta L=2$ decays are crucial to disentangle the Dirac vs Majorana nature of neutrinos. While neutrinoless nuclear double beta decays are the most promising tool to look for di-electron signals of $\Delta L=2$, searches of di-lepton channels with other flavors are far less limited by statistics. We revisit the calculation of $\Delta L=2$ decays of hyperons $B^-_i\to B^+_f\ell^-\ell'^-$. We compute the rates of these decays by including the hyperon form factors, which yields finite results in the one-loop model mechanism involving Majorana neutrinos  (long-range contributions). In addition, we study the short-range contributions to these kinds of processes in two appealing scenarios. First,  we consider the effects of heavy Majorana neutrinos in low-scale seesaw models that involve a minimal parametrization with only two heavy Majorana states. Second, we study an alternative model where $\Delta L=2$ decays are induced by a scalar boson coupled to di-leptons to provide additional predictions.
\end{abstract}

\maketitle

\section{Introduction}

The nature of neutrinos and the origin of their masses are among  the most intriguing questions in particle physics since the discovery of neutrino oscillations. Unlike the charged fermions,  right-hand components of neutrino fields  required to give them an electroweak mass are not protected by chirality, therefore, they can generate Majorana mass terms. If Majorana masses are introduced, the accidental lepton number symmetry in the original formulation of the standard model is explicitly broken by two units  $\Delta \textrm{L}=2$. Therefore, the observation of lepton number violating (LNV) transitions is widely viewed as the cleanest test of the Majorana nature of neutrinos. 

The most extensive and sensitive laboratory to probe LNV is neutrinoless double beta decay (\dbeta) of nuclei.  The amplitude for such transitions is proportional to the ``\emph{effective Majorana neutrino mass}'' $m_{ee}$, which is defined by 
\be
m_{\ell\ell'} \equiv \sum_{j}m_{\nu_j}U_{\ell j}U_{\ell' j}\ .
\label{emmass}
\ee
Here the sum on $j$ runs over all the neutrino mass eigenstates, and $U$ describes the mixing matrix in the leptonic sector.  The non-observation of \dbeta decays in nuclei has set direct strong limits on $m_{ee}$. Currently, the most stringent limit given by the KamLAND-Zen collaboration $T_{1/2}^{0\nu}>1.07\times 10^{26}$ yr ($90\%$ C.L.) on the $^{136}$Xe \dbeta decay half-time  \cite{ParticleDataGroup:2020ssz,KamLAND-Zen:2016pfg} implies the direct limit  $\vert m_{ee} \vert \equiv \vert \sum_i U_{ei} m_i \vert < (61-165)$ meV. Direct bounds on other entries of Eq. (\ref{emmass}) are loose mainly because of the limited statistics of experiments.
From a theoretical point of view, nuclear transitions are limited in precision due to the model-dependent uncertainties of the nuclear matrix elements \cite{Rodejohann:2011mu, Vergados:2012xy, Barea:2013bz, Engel:2016xgb, Dolinski:2019nrj} and are sensitive only to the two electrons channel.  
There is also a large amount of studies related to $\Delta L=2$ processes in decays of tau leptons \cite{Ilakovac:1995wc, Gribanov:2001vv, Helo:2010cw, LopezCastro:2012udb, Kim:2017pra, Yuan:2017xdp, Kim:2017pra, LopezCastro:2012rbs}, pseudoscalar mesons \cite{Cvetic:2010rw,Quintero:2011yh, Yuan:2013yba, Castro:2013jsn, Cvetic:2014nla, Cvetic:2015ura, Cvetic:2015naa, Mandal:2016hpr, Moreno:2016cfz, Cvetic:2016fbv, Cvetic:2017vwl, Cvetic:2017gkt, Yuan:2017uyq, Li:2018pag, Wang:2018bgp, Kim:2019xqj, Cvetic:2020lyh},  and $\Lambda_b$ baryons \cite{Mejia-Guisao:2017nzx, Das:2021prm, Das:2021kzi, Zhang:2021wjj}, mainly motivated by the resonant effect produced by an intermediate  Majorana neutrino and their study in flavor-factories experiments \footnote{We refer the interested reader to  Refs. \cite{Atre:2009rg, Deppisch:2015qwa, Cai:2017mow} for excellent reviews on extensive searches of LNV in colliders.}.

In this work, we are interested in $\Delta Q=\Delta L=2$ decays of hyperons ($B_i^-\to B_f^+\ell^-\ell'^-$, $\ell^{(\prime)}=e$ or $\mu$). Their study is complementary to those in nuclei, but with the advantage that the hadronic matrix elements involved are well known at low momentum transfer \cite{Barbero:2002wm, Barbero:2013fc}. Further, $\Delta L=2$ hyperon decays allow the possibility to study channels involving equal-sign muon and electrons in the final state that are not available in nuclear decays due to limited phase space.  Regarding the experimental side, the large dataset of hyperons pairs produced in charmonium decays that has been and is expected to be accumulated in the BESIII experiment \cite{Li:2016tlt, BESIII:2020iwk} has opened the opportunity to search LNV hyperon decays at sensitivities on the order of $10^{-7}$  \cite{BESIII:2020iwk}. The first upper limit on the LNV decay was reported very recently by BESIII $B(\Sigma^- \to p e^-e^-) < 6.7\times 10^{-5}$. Therefore, for the purposes of comparison, future searches for other channels at BESIII will require the predictions in the framework of reference models of LNV, which is one of the purposes of this paper.

The first model of \dlt decays of hyperons proposed to estimate their branching fractions was reported in \cite{Barbero:2002wm}. The mechanism advocated in Ref. \cite{Barbero:2002wm} considers baryons as the relevant degrees of freedom and involves a loop with hyperons and Majorana neutrinos as intermediate states, as shown in Figure \ref{fig1}. As a first approximation the authors in \cite{Barbero:2002wm} neglected the momentum transfer dependence of the vector and axial  form factors describing the weak vertices, and have kept the  lowest lying hyperons as intermediate states. As a consequence of this approximation, the resulting loop functions exhibit a logarithmic ultraviolet divergence which was regulated using a simple cut-off procedure.

A second approach used by the same authors was based on the MIT bag model  \cite{Barbero:2013fc}. In this case, one starts from the most general dimension-nine  Lagrangian that involves six fermion fields and violates lepton number in two units \cite{Li:2007qx}. This approach requires the computation of the hadronic matrix elements of four quark operators, which can be evaluated, for instance, using the MIT bag model.

These two previous works are associated with different underlying New Physics effects, and the evaluations of their decay amplitudes are also different. The results using the MIT bag model in \cite{Barbero:2013fc} use an effective six-fermion $\Delta$L=2 Lagrangian induced by heavy particles which effects are encoded in the Wilson coefficients. On the other hand, the results of the one-loop mechanism are attributed to light neutrino contributions (long-range effects). According to a numerical estimate in \cite{Barbero:2013fc}, by assuming `reasonable' values for the Wilson coefficients and the New Physics scale of LNV   yields a prediction for the  branching ratio of the $\Sigma^- \to p e^-e^-$ decay in the MIT bag model of $\mathcal{O}(10^{-23})$ which is around ten orders of magnitude larger than its prediction based on the loop model $ \mathcal{O}(10^{-33})$ \cite{Barbero:2002wm}.

Keeping in mind the current and expected experimental searches for \dlt decays of hyperons at BESIII \cite{Li:2016tlt, BESIII:2020iwk}, the goal of the present work is to provide refined estimates for $\Delta$L=2 hyperon decays in the loop model mechanism associated to long-range contributions of light neutrinos.  Our results avoid the undesired divergent behavior of the loop integrals encountered in \cite{Barbero:2002wm} by including the  dependence of the hadronic vector and axial form factors on the momentum transfer. A similar  approach was used in Ref.  \cite{Rein:1989tr} to estimate one of the long-distance contributions to $K^+\to \pi^+\nu\bar{\nu}$ decay. Further, we provide an estimation of the short-range effects associated with two concrete models, namely, heavy Majorana neutrinos contributions in the so-called low-scale seesaw models, and doubly charged Higgs boson $H^{--}$ contributions in the Higgs triplet model. Using current bounds for both scenarios we derive similarly suppressed bounds on the branching fractions.

\section{One-loop mechanism for \dlt decays of hyperons induced by light Majorana neutrinos}

\begin{table}[]
\begin{center}
\begin{tabular}{c c c}
\hline \hline
$\Delta S=0$ & $\Delta S=1$ & $\Delta S=2$  \\
$\Sigma^- \to 	\Sigma^+e^-e^-$& $\Sigma ^-\to p e^-e^-$ & $\Xi ^-\to p e^-e^-$ \\
                               & $\Sigma ^-\to p e^-\mu^-$ & $\Xi ^-\to p e^-\mu^-$ \\
                               & $\Sigma ^-\to p \mu^-\mu^-$ & $\Xi ^-\to p \mu^-\mu^-$ \\
& $\Xi^- \to \Sigma^+  e^-e^-$ &\\
& $\Xi^- \to \Sigma^+  e^-\mu^-$ &\\
\hline \hline
\end{tabular}
\caption{\small 1/2-spin hyperon \dlt decays  allowed by kinematics.}
\label{DeltaL-H-transitions}
\end{center}
\end{table}
\dlt  decays of hyperons occur when two down-type quarks in the initial hyperon convert into two up-type quarks to produce the final hyperon. The different hyperon decay channels can be classified according to their change in strangeness $\Delta S\!=\!0,1,2$ as listed in Table \ref{DeltaL-H-transitions}. These processes violate lepton number in two units, and the most plausible mechanism is the exchange of Majorana neutrinos \cite{Barbero:2002wm, Barbero:2013fc}. 

\begin{figure}[h!]
\begin{center}
\includegraphics[scale=.5]{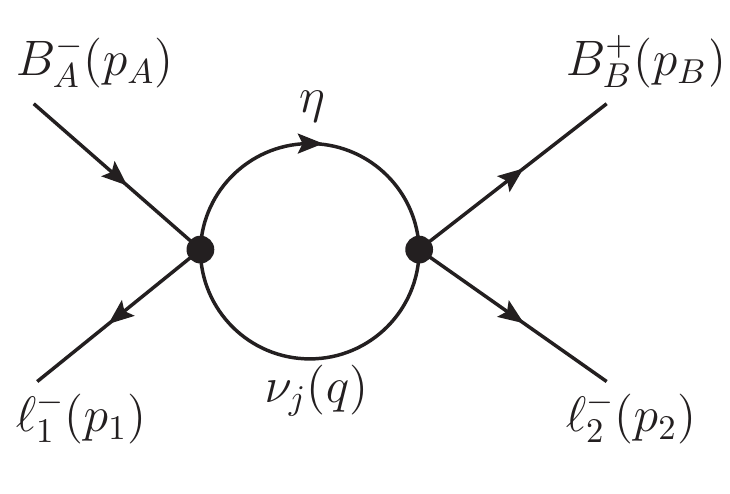}
\end{center}
\caption{\small \dlt decays of hyperons induced by intermediate light Majorana neutrinos in the one-loop model involving an intermediate neutral hyperon $\eta$.}
\label{fig1}
\end{figure}

At the hadron level, this process can be viewed as induced by the one-loop  mechanism shown in Figure \ref{fig1} involving into the loop a neutral hyperon $\eta$ and light Majorana neutrino mass eigenstates $\nu_j$. The corresponding decay amplitude can be written as 

\begin{widetext}
\eq{
i\mathcal{M}=-G^2\sum_j m_{\nu_j}U_{\ell_1 j}U_{\ell_2 j} \sum_{\eta} \int \frac{d^4 q}{(2\pi)^4} \frac{L^{\alpha\beta}_1(p_1,p_2)}{[q^2-m_{\nu_j}^2]}\frac{h_{1\alpha\beta}(p_A,p_B)}{[Q_1^2-m_\eta^2]}-\left[\ell_1(p_1) \leftrightarrow \ell_2(p_2)\right]\ .
\label{amplitude-1}}
\end{widetext}
In the above expression $m_{\nu_j}$ are the masses of Majorana neutrinos and $U_{\ell j}$ their mixings connecting flavor $\ell$ and mass eigenstates. The overall constant $G^2=G_F^2\times (V_{ud}^2, V_{ud}V_{us}, V_{us}^2)$ for $\Delta S\! =\! 0,1,2$, respectively, with $G_F$ the Fermi constant. We have introduced $Q_i=p_A-p_i-q$ as the momentum carried by the hadronic neutral state $\eta$ with the appropriate quantum numbers to contribute as an intermediate state; and  $\ell_1(p_1) \leftrightarrow \ell_2(p_2)$ stands for the contribution of a similar diagram interchanging the final external charged leptons. For the diagram depicted in Figure \ref{fig1}  (see Appendix for the amplitude with interchanged leptons); we have that
\eq{
L^{\alpha\beta}_1(p_1,p_2)&\equiv\bar{u}(p_2)\gamma^{\alpha}(1-\gamma_5)\gamma^{\beta}v(p_1),\\
h_{1\alpha\beta}(p_A,p_B)\equiv&\ \bar{u}(p_B)\gamma_{\alpha}\left[f_{B\eta}(q''^2)+g_{B\eta}(q''^2)\gamma_5\right]\label{Had-par}\\
&(\cancel{Q_1}+m_\eta)\gamma_{\beta}\left[f_{A\eta}(q'^2)+g_{A\eta}(q'^2)\gamma_5\right]u(p_A).\nonumber
}
$f_{(A,B) \eta}$ and $g_{(A,B) \eta}$ denote the vector and axial weak  form factors, respectively, for transitions $A\to \eta$ and $\eta \to B$.  They depend on the squared momentum transfer at each weak vertex, specifically, $f_{A\eta}$ and $g_{A\eta}$ depend on  $q'^2=(p_1+q)^2$, whereas $f_{B\eta}$ and $g_{B\eta}$ depend on $q''^2=(p_2-q)^2$. Their values at zero momentum transfer have been calculated by different groups \cite{Garcia:1985xz,Schlumpf:1994fb,Cabibbo:2003cu} with overall good agreement among them (in Table \ref{t-ff} we quote the values reported in Ref. \cite{Cabibbo:2003cu}).
\begin{table}[]
\begin{center}
\begin{tabular}{c c c c c c}
\hline \hline
Transition  & $\eta$  & $f_{A\eta}$  & $g_{A\eta}$  & $f_{B\eta}$  & $g_{B\eta}$  \\ \hline
$\Sigma^-\to \Sigma^+$  & $\Lambda$  & 0 & 0.656 & 0 & 0.656 \\
 & $\Sigma^0$  & $\sqrt{2}$  & 0.655 & $\sqrt{2}$  & -0.656  \\ 
$\Sigma^-\to p$ & $n$ & -1 & 0.341 & 1 & 1.267 \\ 
 & $\Sigma^0$  & $\sqrt{2}$ & 0.655 & -$1/\sqrt{2}$  & 0.241 \\ 
 & $\Lambda$ & 0 & 0.656 &  -$\sqrt{3/2}$ & -0.895 \\ 
$\Xi^-\to \Sigma^+$ & $\Xi^0$ & -1 & 0.341 & 1 & 1.267 \\
 & $\Sigma^0$  & $1/\sqrt{2}$ & 0.896 & $\sqrt{2}$ & -0.655 \\ 
 & $\Lambda$ & $\sqrt{3/2}$  & 0.239 & 0 & 0.656  \\ 
$\Xi^-\to p$ & $\Sigma^0$ & $1/\sqrt{2}$ & 0.896  & -$1/\sqrt{2}$ & 0.241 \\ 
 & $\Lambda$ & $\sqrt{3/2}$ & 0.239 & -$\sqrt{3/2}$ & -0.895 \\ 
\hline \hline
\end{tabular}
\end{center}
\caption{\small Vector and axial transition form factors for weak hyperon decays at zero momentum transfer. Here $\eta$ stands for the intermediate baryon state, and the subscript $A$ $(B)$ represents the initial (final) baryon \cite{Cabibbo:2003cu}. }
\label{t-ff}
\end{table}

The hadronic part $h_{1\alpha\beta}$ can be rearranged conveniently  as follows 
\eq{
h_{1\alpha\beta}=&\bar{u}(p_B)\gamma_{\alpha}\left[\left(\kappa_{v_+}+ \kappa_{a_+}\gamma_5 \right)\cancel{Q_1}\right. \label{Had-par-2}\\
 &+\left. m_\eta\left( \kappa_{v_-} - \kappa_{a_-} \gamma_5\right) \right]\gamma_{\beta}u(p_A),\nonumber
}where
\eq{
\kappa_{v_\pm}(q^2)&\equiv f_{A\eta}(q'^2)f_{B\eta}(q''^2)\pm g_{A\eta}(q'^2)g_{B\eta}(q''^2),\label{kappas-v} \\ 
\kappa_{a_\pm}(q^2)&\equiv f_{B\eta}(q''^2)g_{A\eta}(q'^2)\pm g_{B\eta}(q''^2)f_{A\eta}(q'^2).\label{kappas-a}
}
It turns out convenient to define 
\eq{
H_{1\alpha\beta}\equiv \sum_{\eta,j} m_{\nu_j}U_{\ell_1 j}U_{\ell_2 j}\int\frac{d^d q}{(2\pi)^d}\frac{h_{1\alpha\beta}}{[q^2-m_{\nu_j}^2][Q_1^2-m_\eta^2]}\ ,\label{integral}
} 
which, after the loop integration, it can be set into the following general form
\begin{widetext}
\eq{H_{1\alpha\beta}=&\sum_{\eta,j} m_{\nu_j}U_{\ell_1 j}U_{\ell_2 j}\bigg\{
\bar{u}(p_B)\gamma_{\alpha}\left[\left(C_{1v_0}^{\eta j}+ C_{1a_0}^{\eta j}\gamma_5 \right)m_{\eta}+ \left(C_{1v_1}^{\eta j}+ C_{1a_1}^{\eta j}\gamma_5 \right)\cancel{p}_1\right. \label{Hpartl}\\ & +\left. \left(C_{1v_2}^{\eta j }+ C_{1a_2}^{\eta j}\gamma_5 \right)\cancel{p}_2+\left(C_{1v_A}^{\eta j}+ C_{1a_A}^{\eta j}\gamma_5 \right)\cancel{p}_A \right]\gamma_{\beta}u(p_A)\bigg\},\nonumber
}
\end{widetext}
where the $C_{1v_{r}}^{\eta j}$ and $C_{1a_r}^{\eta j}$ ($r=0,1,2,A$) functions encode  the effects of the strong interaction relevant in the loop computation. They will depend in general on the neutrino masses and on the two independent Mandelstam variables $t\equiv (p_A-p_1)^2$, and $u\equiv (p_A-p_2)^2$. In this way, the decay amplitude (\ref{amplitude-1}) can be expressed just as the product of the leptonic and hadronic tensor currents as follows
\eq{
i\mathcal{M}=-G^2 \left(L^{\alpha\beta}_1(p_1,p_2)H_{1\alpha\beta} - L^{\alpha\beta}_2(p_1,p_2)H_{2\alpha\beta}\right),
\label{amplitude-2}
}
where the second term in the above expression represents the contribution of the diagram with the final charged leptons interchanged (see Appendix).

\subsection{Hyperon form factors}\label{hff}

The loop integration requires a proper modeling of hyperon form factors in all the range of momentum transfer scales. While SU(3) flavor symmetry considerations are useful  to fix the form factors at zero momentum transfer $(q^2=0)$, we ignore their behavior at finite and large values of $q^2$. From neutrino and electron scattering off nucleons it has been found that the observed distributions can be described by a dipole parametrization. An extrapolation to the timelike region leads to the dipole form factors given by
\eq{
f_i(q^2)&=f_i(0)\left(1-\frac{q^2}{m_{df_i}^2}\right)^{-2},\\ 
g_i(q^2)&=g_i(0)\left(1-\frac{q^2}{m_{dg_i}^2} \right)^{-2},\label{type-pole-aprox}
}
with $ m_{df_i} =0.84$ GeV and $ m_{dg_i} =  1.08$ GeV. Since these pole masses corresponds to strangeness-conserving form factors, a rescaling using  the values of vector and axial mesons masses allows to assume that $ m_{df_i} =0.97$ GeV and $ m_{dg_i} =  1.25$ GeV  would be a good guess for the dipole masses in the strangeness-changing case \cite{Ratcliffe:2004jt}. The values of the form factors at zero momentum transfer, $f_i(0)$ and $g_i(0)$ are given in Table \ref{t-ff} and in the case of the vector form factors they incorporate the effects of SU(3) flavor symmetry breaking \cite{Schlumpf:1994fb, Ratcliffe:2004jt, Mateu:2005wi}.

In Ref.\cite{Barbero:2002wm}, the transition form factors $f_{\{A,B\}_{\eta}}$ and $g_{\{A,B\}_{\eta}}$ were approximated by their values at zero momentum transfers in eqs. (\ref{kappas-v}) and (\ref{kappas-a}). Under such approximation, the relevant $C_{v_r}^{\eta j}$ and $C_{a_r}^{\eta j}$ factors  in eq. (\ref{Hpartl}) are given by
\eq{
C^{\eta j}_{v_0}&=i\frac{\kappa_{v_-}(0)}{16\pi^2}B_0(t,m_{\nu_j}^2,m_\eta^2), \label{v-two-points-functions}\\
C^{\eta j}_{v_A}=-C^{\eta j}_{v_1}&=i\frac{\kappa_{v_+}(0)}{16\pi^2}\left[B_0(t,m_{\nu_j}^2,m_\eta^2)+B_1(t,m_{\nu_j}^2,m_\eta^2)\right],\nonumber\\
C^{\eta j}_{v_2}&=0,\nonumber
} and
\eq{
C^{\eta j}_{a_0}&=-i\frac{\kappa_{a_-}(0)}{16\pi^2}B_0(t,m_{\nu_j}^2,m_\eta^2),\label{a-two-points-functions}\\
C^{\eta j}_{a_A}=-C^{\eta j}_{a_1}&=i\frac{\kappa_{a_+}(0)}{16\pi^2}\left[B_0(t,m_{\nu_j}^2,m_\eta^2)+B_1(t,m_{\nu_j}^2,m_\eta^2)\right],\nonumber\\
C^{\eta j}_{a_2}&=0,\nonumber
}
where $B_0$ and $B_1$ are the two point scalar and vector Passarino-Veltman functions, respectively (see Appendix). Analytical expressions for the loop functions can be derived straightforwardly using Feynman parametrization. We have calculated these expressions and have found good numerical agreement with the results reported  in \cite{Barbero:2002wm} where, however, the mass of the neutrino in the propagator term was neglected. The important point to highlight here is that both $B_0$ and $B_1$ are ultraviolet divergent  and the resulting amplitude for \dlt becomes logarithmically divergent as it was found in Ref. \cite{Barbero:2002wm} using a simple cut-off procedure 

As discussed before, taking the form factors as constants is just an approximation that leads to a divergent amplitude. This bad behavior can be cured by a form factor that vanishes at large $q^2$. The dipole form factors shown in Eq. (\ref{type-pole-aprox}) satisfy the low energy limits dictated by SU(3) flavor symmetry and describe well electron and neutrino scattering data \cite{Garcia:1985xz, Cabibbo:2003cu, Schlumpf:1994fb, Ratcliffe:2004jt, Mateu:2005wi}. In the limit of large momentum transfer, both the vector and axial dipole form factors behave as $\sim 1/q^4$. Considering the dipole approximation would require evaluating an integral with six propagators which, in general, are very difficult to evaluate even numerically. Then, in analogy with the case of meson form factors considered in Ref. \cite{Rein:1989tr} which behave as $\sim 1/q^2$ for large $q^2$, we will use instead a monopolar approximation. The approximation of the dipolar $d$ by the monopolar $m$ form factors is achieved by comparing their slopes at low momentum transfers; this leads  to identify $m_m=m_d/\sqrt{2}$ for the vector and axial poles of the monopolar approximation in the $\Delta S=0,1$ cases.  

Using monopolar expressions for the form factors, the loop integrals become also finite.  The expressions of the relevant $C_{v_r}^{\eta j}$ functions are given by:  
\begin{widetext}
\eq{
C_{v_0}^{\eta j}&=\frac{i}{16\pi^2}\left[ f_{A_{\eta}}(0)f_{B_{\eta}}(0) m_{m_{f_A}}^2 m_{m_{f_B}}^2 D_0 (m_{m_{f_A}},m_{m_{f_{B}}})-g_{{A_{\eta}}}(0)g_{B_{\eta}}(0) m_{m_{g_A}}^2m_{m_{g_B}}^2 D_0(m_{m_{g_A}},m_{m_{g_B}}) \right],\label{v-four-points-functions}\\
C_{v_A}^{\eta j}&=\frac{i}{16\pi^2}\left[f_{A_{\eta}}(0)f_{B_{\eta}}(0) m_{m_{f_A}}^2m_{m_{f_B}}^2 \left[D_1(m_{m_{f_A}},m_{m_{f_{B}}})+D_0(m_{m_{f_A}},m_{m_{f_{B}}})\right]\right. \nonumber\notag \\
&+\left. g_{A_{\eta}}(0)g_{B_{\eta}}(0) m_{m_{g_A}}^2m_{m_{g_B}}^2 \left[D_1(m_{m_{g_A}},m_{m_{g_B}})+D_0(m_{m_{g_A}},m_{m_{g_B}})\right]\right]\nonumber\\
C_{v_1}^{\eta j}&=-\frac{i}{16\pi^2}\left[f_{A_{\eta}}(0)f_{B_{\eta}}(0)m_{m_{f_A}}^2 m_{m_{f_B}}^2 \left[D_2(m_{m_{f_A}},m_{m_{f_{B}}})+D_1(m_{m_{f_A}},m_{m_{f_{B}}})+D_0(m_{m_{f_A}},m_{m_{f_{B}}})\right]\right.\notag \\
&+\left. g_{A_{\eta}}(0)g_{B_{\eta}}(0) m_{m_{g_A}}^2m_{m_{g_B}}^2\left[D_2(m_{m_{g_A}},m_{m_{g_B}})+D_1(m_{m_{g_A}},m_{m_{g_B}})+D_0(m_{m_{g_A}},m_{m_{g_B}})\right]\right],\nonumber\\
C_{v_2}^{\eta j}&=\frac{i}{16\pi^2}\left[f_{A_{\eta}}(0)f_{B_{\eta}}(0) m_{m_{f_A}}^2m_{m_{f_B}}^2 D_3(m_{m{f_A}},m_{m_{f_{B}}})+g_{A_{\eta}}(0)g_{B_{\eta}}(0)m_{m_{g_A}}^2m_{m_{g_B}}^2 D_3(m_{m_{g_A}},m_{m_{g_B}})\right], \nonumber
}
where (in GeV units) $m_{m_{fA}}=0.84\, (0.97)/\sqrt{2}$ and  $m_{m_{gA}}=1.08\, (1.25)/\sqrt{2}$ for $\Delta S=0\, (1)$ transitions, as explained previously. The Passarino-Veltman functions are given by 
\eq{
D_{\{0,1,2,3\}}(m_{m_{f_A}},\,m_{m_{f_B}})&\equiv D_{\{0,1,2,3\}} \left(t,m_A^2,s,m_2^2,m_1^2,m_B^2,m_{\nu_j},m_\eta,m_{m_{f_A}},m_{m_{f_B}} \right),\label{DPAVE}\\
D_{\{0,1,2,3\}}(m_{m_{g_A}},\,m_{m_{g_B}})&\equiv D_{\{0,1,2,3\}} \left(t,m_A^2,s,m_2^2,m_1^2,m_B^2,m_{\nu_j},m_\eta,m_{m_{g_A}},m_{m_{g_B}} \right),\nonumber
}
\end{widetext}
and  $s=(p_1+p_2)^2=m_A^2+m_B^2+m_1^2+m_2^2-t-u$ is the other Maldemstam variable. The functions  $C_{a_r}^{\eta j}$ can be obtained straightforwardly from the above expressions considering the following replacements 
\eq{
C_{a_0}^{\eta j}&= -C_{v_0}^{\eta j}\left(f_{A\eta}\leftrightarrow g_{A\eta},\  m_{m_{f_A}} \leftrightarrow m_{m_{g_A}}\right),\label{a-four-points-functions}\\
C_{\{a_1,\,a_2,\, a_A \}}^{\eta j}&= C_{\{v_1,\,v_2,\, v_A \}}^{\eta j}\left(f_{A\eta}\leftrightarrow g_{A\eta},\  m_{m_{f_A}} \leftrightarrow m_{m_{g_A}}\right).\nonumber
 }
Given that the four-point functions arising from the monopolar form factors are not divergent, the amplitudes are finite and physical.

\subsection{Numerical analysis (one-loop mechanism)}

\begin{figure}
\begin{center}
\begin{tabular}{cc}
\includegraphics[scale=.43]{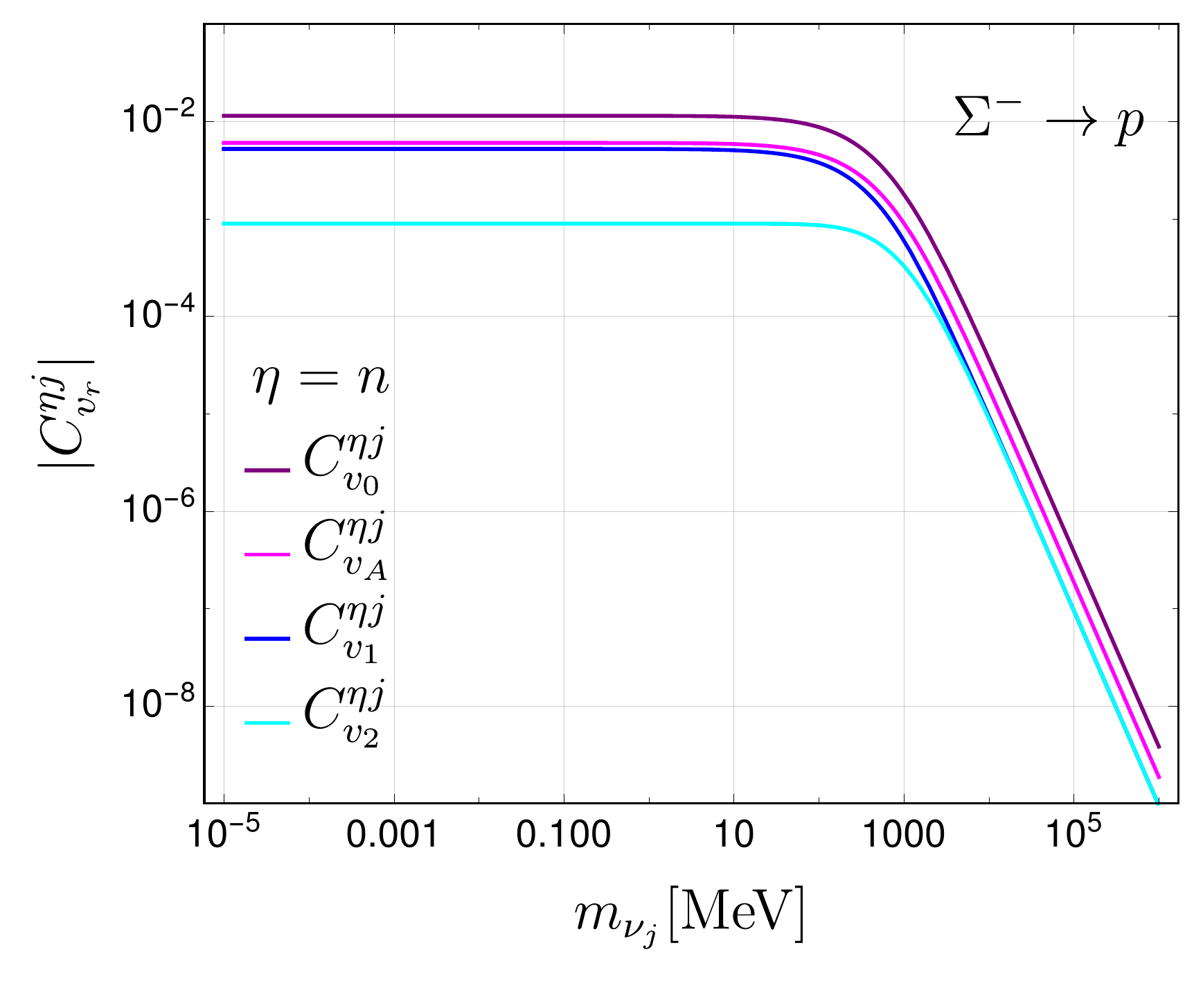}&
\end{tabular}
\end{center}
\caption{ Individual $C_{v_r}^{n j}$ loop-factors as function of the neutrino mass in the monopole form factors model for the decay chain $\Sigma^-\to n^* \to p$. For illustration purposes, we have used the maximum  values for the Lorentz invariants $(t,s)$ that are allowed by kinematics, as well as $m_1=m_2=m_e$ in eqs. (\ref{v-four-points-functions}) and (\ref{DPAVE})}.
\label{ff-plot}
\end{figure}
The hadronic matrix element defined  in eqs. (\ref{Hpartl}) and (\ref{amplitude-2}) depends on the effective total form factors $C_{v_r,a_r}^{\eta}$; for each intermediate hadronic state $\eta$ in the loop, they can be written as
\eq{
C_{v_r}^{\eta}&\equiv \sum_j m_{\nu_j}U_{\ell_1 j}U_{\ell_2 j}C_{v_r}^{\eta j},\quad (v_r=v_0,v_1,v_2,v_A),\label{relevant-factors}\\
C_{a_r}^{\eta}&\equiv \sum_j m_{\nu_j}U_{\ell_1 j}U_{\ell_2 j}C_{a_r}^{\eta j},\quad (a_r=a_0,a_1,a_2,a_A),\nonumber
}
where the individual $C_{v_r, a_r}^{\eta j}$ factors, determined from the loop integration, depend in general upon the neutrino mass $m_{\nu_j}$ involved in the neutrino propagator  (\ref{integral}).  Figure \ref{ff-plot} shows the absolute value of the $C_{v_r}^{\eta j}$ (with $\eta=n$) form factors as a function of the intermediate neutrino mass $m_{\nu_j}$ for the specific $\Sigma^-\to p$ transition, and using the monopolar approximation described in  equations (\ref{v-four-points-functions})  (similar results are obtained for the rest of the decay channels listed in Table \ref{t-ff} as well as for the analysis of the axial $C_{a_r}^{\eta j}$ form factors). From this plot we observe that the dominant contribution arises from the $C_{v_0}^{\eta j}$ coefficient. Also, for light neutrinos ($ m_{\nu_j} \lesssim 100$ MeV), all the $C_{v_r}^{\eta j}$ factors are insensitive to the neutrino mass value. However, for heavier neutrino states, the $C_{v_r}^{\eta j}$ one-loop functions describing the \dlt hyperon decays become strongly dependent on the neutrino mass. When $m_{\nu}\sim 100$ MeV, the loop integral becomes sensitive to the neutrino mass because this scale starts to be comparable to the involved hadronic scales (pole and baryon masses in the loop). Formally, one expects that particles much heavier than the ones included as explicit degrees of freedom in these loop calculations should be considered as short-distance contributions. This means that this loop mechanism is not valid for particles much heavier than a few hundreds of MeV (see below for the case of heavy Majorana neutrinos).

On the other hand, if we assume that only very light neutrino states exist, the effective form factors in eq. (\ref{relevant-factors}) can be approximated  by
\eq{
C_{v_r}^{\eta}& \equiv  m_{\ell_1\ell_2}  C_{v_r}^{\eta 0},\label{relevant-factors-A}\quad
C_{a_r}^{\eta} \equiv  m_{\ell_1\ell_2}  C_{a_r}^{\eta 0},
}where $ m_{\ell_1\ell_2} $ is the effective Majorana mass parameter, and $C_{v_r}^{\eta 0}$ ($C_{a_r}^{\eta 0}$) are the one-loop functions in eq. (\ref{DPAVE}) evaluated at $m_{\nu_j}=0$~\footnote{In Ref. \cite{Barbero:2002wm} the masses of  neutrinos in the loop integral are set to zero from the beginning. In contrast, our results are rather general and appropiate to evaluate the effects of new states until $1$ GeV (see discussion in the main text).}. We note that in Ref. \cite{Barbero:2002wm} the values $m_{ee}=10$ eV, and $m_{\mu\mu}=10$ MeV were used as arbitrary inputs for the normalization of the decay rates; in addition, that reference did not include the $\mu e$ decay channels as we do in the present calculation. Note that the direct upper limits for $m_{\ell\ell'}$ reported in \cite{KamLAND-Zen:2016pfg,Atre:2005eb} are given by \footnote{A recent work presented in \cite{Fuks:2020zbm} proposes that the study of the non-resonant signature $pp\to\ell^{\pm}\ell^{'\pm}$ at the LHC can be used to test $m_{\mu\mu}$ to a sensitivity of $\vert m_{\mu\mu}\vert\sim 7.3$ GeV.}:
\eq{
\vert m_{ee}\vert &<0.165\, \textrm{eV},\label{LIMMAJ}\\
\vert m_{e\mu}\vert &< 90\, \textrm{GeV},\nonumber\\
\vert m_{\mu\mu} \vert &<480\, \textrm{GeV}.\nonumber\
}

By computing numerically the form factors in Eq. (\ref{relevant-factors-A}), we have obtained the branching ratios listed in Table \ref{BRs}. We observe that channels involving two electrons are strongly suppressed due to the strong limits imposed from \dbeta nuclear decay. 
On the other hand, by assuming the direct upper limits in Eq. (\ref{LIMMAJ}) for the $e\mu$ and $\mu\mu$ effective masses, we would get $\textrm{BR}(\Sigma^-\to p\mu\mu)=1.7\times 10^{-10}$ and $\textrm{BR}(\Sigma^-\to pe\mu)=1.6\times 10^{-12}$ which appear to be close to the projected sensitivity of BES-III \footnote{In this `naive' approximation one also gets $\textrm{BR}(\Xi^-\to \Sigma^+\mu e)=1.8\times 10^{-14}$, $\textrm{BR}(\Xi^-\to p\mu \mu)=2.5\times 10^{-11}$, and $\textrm{BR}(\Xi^-\to p\mu e)=2.3\times 10^{-12}$, which also look closer to the experimental sentitivity of BES-III.}. These large ratios should be taken with care because the upper limits used for $m_{e\mu}$ and $m_{\mu\mu}$ lie beyond the range of validity of this scenario, according to Figure \ref{ff-plot}.   If we assume the maximal value $|m_{e\mu}|=|m_{\mu\mu}|=100$ MeV consistent with the approximation in Eq. (\ref{relevant-factors-A}) in the loop-model mechanism, then we obtain the rates reported in Table (\ref{BRs}). 
\begin{table}[]
\begin{center}
\begin{tabular}{l c r}
Transition & & Branching Ratio \\ \hline \hline
& &\textit{One-loop model} \\
\hline \hline
$\Sigma^- \to \Sigma^+ee$  &  & 1.6\,(0.4)\,\,$\times\, 10^{-41}$  \\ \hline
$\Sigma^- \to  p ee$      &    & 2.2\,(0.4) $\times\, 10^{-34}$ \\ \hline
$\Sigma^- \to  p\mu\mu$     &      & 7.4\,(1.6) $\times\, 10^{-18}$   \\ \hline
$\Sigma^- \to p\mu e$       &     & 7.2\,(1.6) $\times\, 10^{-17}$  \\ \hline
$\Xi^-    \to \Sigma^+ee$   &   & 2.0\,(0.7) $\times\, 10^{-36}$  \\ \hline
$\Xi^-  \to  \Sigma^+\mu e$ &  & 2.8\,(0.9) $\times\, 10^{-20}$ \\ \hline
$\Xi^-    \to  pee$         &    & 7.1\,(2.0) $\times\, 10^{-36}$ \\ \hline
$\Xi^-    \to p\mu\mu$      &   & 1.1\,(0.4) $\times\, 10^{-18}$  \\ \hline
$\Xi^-    \to  p\mu e$       &   & 3.7\,(1.2) $\times\, 10^{-18}$  \\ \hline
\hline
\end{tabular}
\caption{\small Branching ratios of $\Delta L=2$ hyperon decays induced by Majorana neutrinos in the one-loop model mechanism.  We consider the upper limit of the effective Majorana mass $|m_{ee}|$ as given in eq. (\ref{LIMMAJ}), but we assume $|m_{e\mu, \mu\mu}|_{\rm max}\sim 100$ MeV  that is consistent with the approximation in Eq. (\ref{relevant-factors-A}) (see also Figure \ref{ff-plot}). Quoted errors (within parentheses) are estimated by varying the pole mass $m_m$ by $\pm 15\%$.
}
\label{BRs}
\end{center}
\end{table}

\section{Short-range contributions}

If LNV is mediated by heavy particles, then an appropriate framework to deal with such effects corresponds to an effective field theory analysis \cite{
Gonzalez:2015ady, Arbelaez:2016uto, Geib:2016atx, Quintero:2016iwi, Geib:2016daa, Liao:2019gex, Liao:2021qfj}. In this regard, we will consider the most general six-fermion effective interaction describing $\Delta$L=2 processes involving any leptonic and hadronic state with second and/or third generation of quarks \cite{Gonzalez:2015ady, Arbelaez:2016uto, Geib:2016atx, Quintero:2016iwi, Geib:2016daa, Liao:2019gex, Liao:2021qfj}, following the notation in \cite{Quintero:2016iwi} this can be written as follows
\eq{
\mathcal{L}^{\Delta \textrm{L}=2}_{\textrm{eff}}=\frac{G_F^2}{\Lambda}\sum_{i,X,Y,Z}\left[C_i^{XYZ} \right]_{\alpha \beta} \mathcal{O}_i^{XYZ},
\label{efl}
}
where $C_i$ are effective dimensionless couplings, and $\Lambda$ is the heavy mass scale of New Physics. The dimension-9 operators are classified by
\eq{
\mathcal{O}_1^{XYZ}&=4[\bar{u}_iP_X d_k][\bar{u}_jP_Y d_n](j_Z),\label{op1}\\
\mathcal{O}_2^{XYZ}&=4[\bar{u}_i\sigma^{\mu\nu}P_X d_k][\bar{u}_j\sigma_{\mu\nu}P_Y d_n](j_Z),\nonumber\\
\mathcal{O}_3^{XYZ}&=4[\bar{u}_i\gamma^{\mu}P_X d_k][\bar{u}_j\gamma_{\mu}P_Y d_n](j_Z),\nonumber\\
\mathcal{O}_4^{XYZ}&=4[\bar{u}_i\gamma^{\mu}P_X d_k][\bar{u}_j\sigma_{\mu\nu}P_Y d_n](j_Z)^{\nu},\nonumber\\
\mathcal{O}_5^{XYZ}&=4[\bar{u}_i\gamma^{\mu}P_X d_k][\bar{u}_jP_Y d_n](j_{Z})_\mu,\nonumber
}and the leptonic currents are defined as 
\eq{
j_Z&=\bar{\ell}_\alpha P_Z\ell_\beta^c,\quad j_Z^{\nu}=\bar{\ell}_\alpha\gamma^{\nu}P_Z\ell_\beta^c.
} 
In the above expressions $P_{X,Y,Z}$ ($X,Y,Z=L,\textrm{ or } R$) are the left and right proyectors $P_{L,R}=1/2(1\mp \gamma_5)$, whereas $\alpha,\beta$ denote one of three lepton flavors ($e, \mu, \tau$). Then, assuming that short-range contributions are the dominant ones, the decay amplitude of $\Delta$L=2 hyperon decays is given by
\eq{
\mathcal{M}(B_A^-\to B_B^+\ell_1\ell_2)&=\langle B_B^+ \ell_1\ell_2\vert \mathcal{L}^{\Delta \textrm{L}=2}_{\textrm{eff}}\vert B_A^-\rangle,\label{Amp-efec} \\
&=\frac{G_F^2}{\Lambda}\sum_i [C_i^{X,Y,Z}]_{\ell_1\ell_2}\mathcal{F}_i,\nonumber
}where the $\mathcal{F}_i$ functions describe the matrix elements associated to the all different operators in eq. (\ref{op1}).
Below we provide two concrete UV completions of the local six-fermion effective Lagrangian described by the eq. (\ref{efl}).

\subsection{Heavy neutrino contributions}

\begin{figure}[h!]
\begin{center}
\begin{tabular}{cc}
\includegraphics[scale=.45]{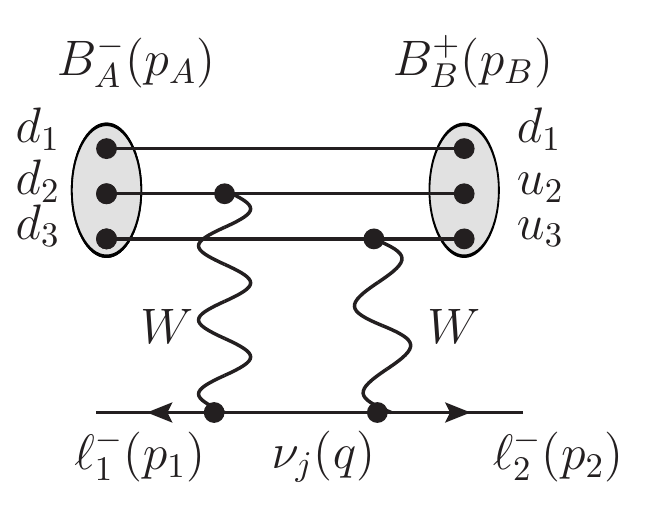}\quad \quad&\includegraphics[scale=.4]{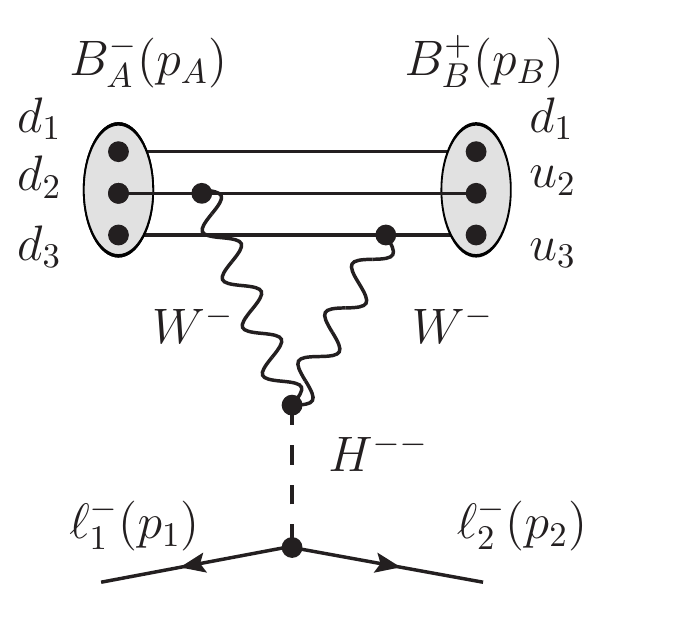}\\
{\small (a) }& {\small(b)}
\end{tabular}
\caption{Two possible UV completions for the effective lagrangian in eq. (\ref{efl}). (a) Heavy neutrino contributions from low-scale seesaw models.  (b) Doubly charged scalar contributions in the Higgs triplet model (HTM).}\label{short-range}
\end{center}
\end{figure}
We evaluate first the contribution of heavy neutrinos that can appear in many low-scale seesaw models, for that, we will consider the minimal parametrization presented in reference \cite{Ilakovac:1994kj,Hernandez-Tome:2019lkb}.
In this model, the neutrino sector contains 5 Majorana fields ($\chi_i=\chi_{L_i}+\chi_{L_i}^c$); after mass matrix diagonalization one ends with three light (active) neutrinos that determine the observed oscillation  phenomena and two heavy neutrinos $N_{1,2}=\chi_{4,5}$ states. The charged weak lepton current relevant for our computation is described by the lagrangian:
\eq{
\mathcal{L}_{W}^{\pm}=-\frac{g}{2\sqrt{2}}W_{\mu}^-\sum_{i=1}^{3}\sum_{j=1}^{5}B_{ij}\bar{\ell}_i\gamma^{\mu}(1-\gamma_5)\chi_j+\textrm{h.c.},\label{charged-currents}
}where $B$ is a $3\times 5$  matrix
\eq{
B_{ij}=\sum_{k=1}^3\delta_{ik}U^{\nu}_{k_j},\label{Bmat}
}and $U$ is the matrix that diagonalizes the neutrino mass matrix. The relevant point for our discussion is that all the genuine effects of LNV due to presence of heavy $N_{1,2}$ neutrino masses can be parametrized in terms of their mass splitting ($r=m_{N_2}^2/m_{N_1}^2$, with $r\not = 1$) which determine their Majorana characteristics. Note that, in this scenario, when the two Majorana states are degenerate they form a Dirac singlet neutrino and lepton number is exactly conserved. Then the elements of the matrix $B$ involving the heavy states in eq. (\ref{Bmat}) can be expressed in terms of the heavy-light mixings $s_{\nu_\alpha}$ (with $\alpha,\beta=e,\mu, \tau$) and the $r$ parameter as follows \cite{Ilakovac:1994kj}
\eq{
B_{\alpha N_1}=-i\frac{r^{1/4}}{\sqrt{1+r^{1/2}}}s_{\nu_\alpha},\quad B_{\alpha N_2}=\frac{1}{\sqrt{1+r^{1/2}}}s_{\nu_\alpha}.
} 
For this parametrization, the contribution of heavy neutrinos to  $\Delta$L=2 hyperon  decays is given by
\eq{
\mathcal{M}&=-G^2 \sum_{j=1}^{2} \frac{B_{\ell_1 N_j}B_{\ell_2 N_j}}{2m_{N_j}}\, \textrm{X}^{\mu\nu}L_{\mu\nu},\label{amp-HB}
} where
\eq{
\textrm{X}^{\mu\nu}\equiv& \langle B_B^-(p_B) \vert \Gamma^{\mu\nu} \vert B_A^+(p_A)\rangle,\label{V-A-had}\\
\Gamma^{\mu\nu}=&\left[ \bar{u}\gamma_\mu (1-\gamma_5)D\right]\times\left [\bar{u}\gamma_\nu (1-\gamma_5)D'\right]\nonumber.
}
is the hadronic matrix element describing the transition from $B_A^-$  to $B^+_B$ ($D$  and $D'$ stands for down-type quarks $d$ or $s$ according to the initial and final states). Because the transition between quarks of initial and final states is via the weak charged current, as depicted in Fig. \ref{short-range}(a),  the tensor hadronic current is given by the product of two bilinear $V-A$ structures. On the other hand, the leptonic part can be simplified as follows
\eq{
L_{\mu\nu}&\equiv \bar{u}(p_{\ell_2})\gamma_\mu(1-\gamma_5)\gamma_\nu v(p_{\ell_1})-\ell_1(p_{\ell_1}) \leftrightarrow \ell_2(p_{\ell_2}),\nonumber\\
&=2g_{\mu\nu}\bar{u}(p_{\ell_2})(1+\gamma_5) v(p_{\ell_1}),\nonumber\\
&\equiv 4g_{\mu\nu}(j_R^{\ell_2\ell_1}). 
}Fortunately, some of the hadronic matrix elements in eq. (\ref{V-A-had}) have been computed in the framework of the so-called MIT bag model in Ref. \cite{Barbero:2013fc}. In the non-relativistic approximation, these hadronic matrix elements involving fourth-quarks operators can be expressed in terms of only two $A$ and $B$ functions, in such a way that after the contraction of Lorentz indices, we have that 
\eq{
X^{\mu\nu}g_{\mu\nu}=\bar{u}(p_B)[A+B\gamma_5]u(p_A),
}where $u(p_A)$ and $u(p_B)$ denotes the spinors of initial and final hadronic states, respectively. Thus, eq. (\ref{amp-HB}) becomes
\eq{
\mathcal{M}=&-2G^2 \sum_{j=1}^2 \frac{B_{\ell_1 N_j}B_{\ell_2 N_j}}{m_{N_j}} \bar{u}(p_B)[A+B\gamma_5]u(p_A)\, (j_R^{\ell_2\ell_1}). \label{HNc}}
Matching eq. (\ref{Amp-efec}) for $\Lambda=m_{N_1}$ with eq. (\ref{HNc})  we obtained the particular realization of the relevant Wilson coefficient
\eq{[C_3^{LLR}]_{\ell_1\ell_2}&=-2V_{uD}V_{uD'}\sum_{j=1}^{2}B_{\ell_1 N_j}B_{\ell_2 N_j},\label{WCHN}\\
&=2V_{uD}V_{uD'}s_{\nu_{\ell_1}}s_{\nu_{\ell_2}}\frac{(r-1)}{(r+r^{1/2})}.\nonumber
}
From the above expressions it turns out clear that when the two heavy states are degenerate (singlet Dirac case) lepton number is exactly conserved as it is expected.

Now, in order to estimate the contributions of heavy neutrinos, we consider as a benchmark the mass-independent indirect limits for the relevant heavy-light mixing angles coming from the latest global fits to electroweak precision observables given in \cite{Fernandez-Martinez:2016lgt, Coutinho:2019aiy}:
\eq{
s_{\nu_e}<0.050, \quad s_{\nu_\mu}<0.021, \quad s_{\nu_\tau}<0.075.\label{INDLIM}
}
We also have to consider maximum perturbative values for masses and mixings of the new states. In this parametrization, such condition translates into the following relation  \cite{Ilakovac:1994kj, Hernandez-Tome:2019lkb}
\eq{
m_{N_1}r^{1/4}<\frac{\sqrt{2}\pi v}{\textrm{max}\{s_{\nu_i}\}}\label{PerCon}.
}
Therefore, assuming the maximal values in (\ref{INDLIM}), the perturbative condition reduces to $m_{N_1}r^{1/4}<8.2\,$TeV.

Taking into account the above constraints we can estimate the rates of \dlt hyperon decays induced by heavy neutrinos in low-scale seesaw models. Let us consider the specific example of the $\Sigma^-\to p\ell\ell'$ decays for which the values  $A=3.56\times 10^5$ MeV$^3$ and $B=0$  were obtained using the eigenfunctions of quarks confined within a baryon in the MIT bag model \cite{Barbero:2013fc}.

Considering maximal values for the heavy-light mixings in eq. (\ref{INDLIM}) and the representative values $M_{N_1}=1 \textrm{ TeV}, r=0.01$ for the masses of the new heavy states consistent with the perturbative limit, we have that  (see appendix for details)
\eq{
BR(\Sigma^-\to pee)&=[C_3^{LLR}]_{ee}^2\cdot(5.0\times 10^{-14}), \label{BRHN}\\ 
&=4.9\times 10^{-30},\nonumber\\
BR(\Sigma^-\to pe\mu)&=[C_3^{LLR}]_{e\mu}^2\,\cdot (4.5\times 10^{-14}), \nonumber\\ 
&=7.0\times 10^{-31},\nonumber\\
BR(\Sigma^-\to p\mu\mu)&=[C_3^{LLR}]_{\mu\mu}^2\cdot ( 4.5\times 10^{-15}), \nonumber\\
&=1.1\times 10^{-32}.\nonumber
}
Bounds for the rest  of hyperon decays can be computed in a similar way if matrix elements of four-quark operators for other channels become available.

\subsection{Higgs triplet model contributions}

Majorana neutrinos are the most appealing but not unique mechanism to generate \dlt transitions in hyperons. As an alternative, we explore the possible effects  that can arise in the presence of doubly charged scalar bosons coupled to dileptons, particularly, in the so-called Higgs Triplet Model (HTM) \cite{Schechter:1980gr}.  Here, the scalar sector is extended by including a complex $SU(2)_L$ Higgs triplet $\Delta$ with $Y=2$ along with the SM doublet $\Phi$. The scalar triplet is parametrized by a $2\times 2$ matrix as follows
\eq{
\Delta=\begin{pmatrix}
\frac{1}{\sqrt{2}}\Delta^+ & \Delta^{++}\\
\Delta^0 & -\frac{1}{\sqrt{2}}\Delta^+
\end{pmatrix},
} 
and the relevant Yukawa lagrangian is given by
\eq{
\mathcal{L}_Y=h_{ij}\psi_{i}^TCi\sigma_2\Delta \psi_{j}+\textrm{H.c},
}where $C$ is the charge conjugation matrix, $\psi_{i}^T=(\nu_{iL}^T,e_{iL}^T)$, $\sigma_2$ is the second Pauli matrix, and $h_{ij}$ are the entries of a $3\times 3$ symmetric Yukawa matrix. 
The neutral component of $\Delta$ developes a vacuum expectation value
$v_\Delta$,  as a consequence, neutrino masses are generated. 

After the spontaneous symmetry breaking the physical scalar spectrum is composed of seven states: two CP-even scalars $H_1$ and $H_2$, one CP-odd scalar $A$, two charged scalars $H^{\pm}$, and two doubly charged scalars $H^{\pm\pm}$ (the  $H^{\pm\pm}$ are completely built out of the triplet fields  $H^{\pm\pm}=\Delta^{\pm\pm})$. 
For the purposes of this work, we only focus on the phenomenology of the doubly charged states;  a complete list of all the new vertices in the HTM can be found in \cite{FileviezPerez:2008jbu}. The coupling for the interaction of the doubly charged scalar $H^{\pm\pm}$ with a pair of $W^\pm$ gauge bosons needed to built the amplitude in Fig. \ref{short-range}(b) is given by: $i\sqrt{2}g^2v_{\Delta}g_{\mu\nu}$. Note that similar contributions replacing each of the weak $W^-$ bosons with a singly charged $H^-$ scalar are suppressed due to small couplings $H^-q_d\bar{q}_u$ proportional to the light quark masses. 

The contribution of the HTM to $\Delta$L=2  hyperon decays is given by 
\eq{
\mathcal{M}=&-4 \sqrt{2}G^2 \frac{h_{\ell_1\ell_2} v_\Delta}{M_{H^{\pm\pm}}^2}X^{\mu\nu}g_{\mu\nu}\,\bar{u}(p_2)(1-\gamma_5)v(p_1),\label{amp-doubly}\\
\equiv&-8\sqrt{2}G^2\frac{h_{\ell_1\ell_2}v_\Delta}{M_{H^{\pm\pm}}^2}\bar{u}(p_B)\left[A+B\gamma_5 \right]u(p_A)\, (j_L^{\ell_2\ell_1}).\nonumber
}Comparing the above expression with eq. (\ref{Amp-efec}) for $\Lambda=M_{H^{\pm\pm}}^2/v_{\Delta}$, the Wilson coefficient associated to the HTM contribution is given by \footnote{In eq. (\ref{amp-doubly}) we have added the diagram with the external charged leptons interchanged using that $\bar{u}(p_2)P_{\{L,R\}}v(p_1)=-\bar{u}(p_1)P_{\{L,R\}}v(p_2)$.}
\eq{
[C_3^{LLL}]_{\ell_1\ell_2}&= -8\sqrt{2}V_{uD}V_{uD'}h_{\ell_1\ell_2}.\label{WCHT}
} 
In order to evaluate this contribution we must consider that $v_\Delta$ is constrained from the correction to the $\rho$ parameter, which after the introduction of the Higgs triplet becomes   
\eq{
\rho=M_{W}^2/M_{Z}^2\cos^2 \theta_W=\frac{1+2v_\Delta^2/v^2}{1+4v_\Delta^2/v^2},\label{rho-cor}
}where $v=246$ GeV is the v.e.v. of the SM doublet. Then, considering the experimental value $\rho^{\rm exp}=1.00038(20) $ \cite{ParticleDataGroup:2020ssz} one is lead to the upper limit $v_\Delta\lesssim \mathcal{O} (1)$ GeV \cite{Akeroyd:2009nu, Dev:2017ouk}. 

Furthermore, $M_{H}^{\pm\pm}$ is constrained indirectly as a function of the product of leptonic Yukawa couplings from several processes \cite{Swartz:1989qz, OPAL:2003kya, BhupalDev:2018vpr,Akeroyd:2009nu, Dev:2017ouk, Quintero:2012jy}, including Bhabha scattering, LFV violating transitions, muonic oscillation, and the electron and muon $(g-2)$ observables (see Table \ref{LimitshM}) \footnote{A recent study of the production of doubly charged Higgs bosons at LHC can be found in \cite{Fuks:2019clu}.}. 

Taking for simplicity non $\tau$-flavored interactions, that is, $h_{\tau i}=0$ ($i=e,\mu,\tau$) and the contrainsts from Table \ref{LimitshM}. We adopt a conservative benchmark considering that $v_\Delta=3$ GeV, and $h_{mm}\simeq 0.1$ ($m=e,\,\mu$) for the rest of diagonal Yukawa couplings. If we now consider the limits from $\ell\ell\to\ell\ell$ ($\ell=e,\mu$) data which only involve diagonal couplings $h_{ee}$ and $h_{\mu\mu}$, then $m_{H^{\pm\pm}}\gtrsim 395$ GeV.
Choosing this lowest value for $m_{H^{\pm\pm}}$, we obtain \footnote{ The branching ratio for the HTM contribution can be obtained directly from eq. (\ref{BRHN}) replacing $[C_3^{LLR}]_{\ell_1\ell_2}$ by $[C_3^{LLL}]_{\ell_1\ell_2}$.}
\eq{
BR(\Sigma^-\to pee)_{HTM}&=1.1\times 10^{-30},\label{ulbdsigma}\\
BR(\Sigma^-\to pe\mu)_{HTM}&=1.3\times 10^{-39},\nonumber\\
BR(\Sigma^-\to p\mu\mu)_{HTM}&=1.0\times 10^{-31}.\nonumber
}
Note that for smaller Yukawa couplings $h_{mm}$, the above upper limit increases by a factor $1/h_{mm}^2$ if we still assume the lower bound on $m_{H^{\pm\pm}}$ from $ee\to\mu\mu$ data quoted in Table (\ref{LimitshM}). Moreover, in this case, it is necessary to consider $h_{e\mu}\lesssim 3.5 \times 10^{-6}$ to obey the strongly constraint coming from $\mu\to eee^+$, as a consequence, the prediction for the ($e\mu$) channel would be more suppressed.

\begin{widetext}

\begin{center}
\begin{table}[h!]
\begin{center}
\begin{tabular}{c c c}
Process & Current data & Constraint [GeV$^{-2}$] \\ \hline \hline
$\mu^-\to eee^+\quad$   & $< 1.0\times 10^{-12}\quad$  & $|h_{ee}^\dagger\, h_{e\mu}|/{M_{H^{\pm\pm}}^2}<2.3\times 10^{-12}$ \\
\hline
$\mu\to e\gamma\quad$   & $< 4.2\times 10^{-13}\quad$  & $ \sum\limits_{k=e,\mu,\tau} \vert  h_{ek}^\dagger\, h_{\mu k}|/{M_{H^{\pm\pm}}^2}<2.7\times 10^{-10} $ \\ \hline
electron $g-2\quad$   & $< 5.2\times 10^{-13}\quad$  & $\sum\limits_{k=e,\mu,\tau} | h_{ek}|^2/{M_{H^{\pm\pm}}^2}<1.2\times 10^{-4} $ \\
\hline
muon $g-2\quad$   & $< 4.0\times 10^{-9}\quad$  & $\sum\limits_{k=e,\mu,\tau} | h_{\mu k}|^2/{M_{H^{\pm\pm}}^2}<1.7\times 10^{-5} $ \\
\hline
muonic oscillation   & $< 8.2\times 10^{-11}$  & $| h_{ee}^{\dagger}h_{\mu\mu}|^2/{M_{H^{\pm\pm}}^2}<1.2\times 10^{-7} $ \\
\hline
$ee\to ee$ (LEP)   & $\Lambda_{\rm eff}> 5.2$ TeV  & $| h_{ee}|^2/{M_{H^{\pm\pm}}^2}<1.2\times 10^{-7} $ \\
\hline
$ee\to \mu\mu$ (LEP)   & $\Lambda_{\rm eff}>7.0$ TeV  & $| h_{\mu\mu}|^2/{M_{H^{\pm\pm}}^2}<6.4\times 10^{-8} $ \\
\hline
\end{tabular}
\caption{\small Current experimental limits for ($e$, $\mu$) flavor processes that constrain the product of Yukawa couplings ($h^{\dagger}h$) as a function of the mass $m_{H^{\pm\pm}}$. A comprehensive and detailed analysis of all the limits including $\tau$ flavor transitions can be found in  \cite{BhupalDev:2018vpr}.}
\label{LimitshM}
\end{center}
\end{table}
\end{center}
\end{widetext}

\section{Conclusions}
We have studied all the $\Delta L=2$ decays of spin-1/2 hyperons $B_A^- \to B_B^+\ell^-\ell'^{-}$   that are allowed by kinematics within a model involving a one loop mechanism with baryons and light Majorana neutrinos as intermediate states. This study improves previous estimates reported in \cite{Barbero:2002wm} in several ways. First, we have included the momentum dependence of hyperons form factors using a monopolar model  which allows to cure the  bad ultraviolet  behavior encountered in \cite{Barbero:2002wm}. Second, because we kept finite values for the Majorana neutrino masses in the loop computation we realize that for  $ m_\nu\gtrsim 100$ MeV, the dependence on the neutrino mass of the relevant loop functions that dictate the strength of $\Delta$L=2 hyperon decays becomes relevant. This indicates that the validity of the loop mechanism for these long-distance contributions can not be extended for neutrino states beyond a few hundreds of MeV.
Except for the overall effective Majorana mass factor $m_{\ell\ell'}$, our calculation basically confirms the strong suppression found in previous results for light neutrinos (loop mechanism). In general, the momentum dependence of the form factors yields results for the branching fraction that are suppressed by three orders of magnitude with respect to the case where this momentum dependence is neglected (Ref. \cite{Barbero:2002wm}). Given the poor current limits on $m_{\mu\mu}$, obtained from $K^+\to \pi^-\mu^+\mu^+$ decays, the bounds obtained for the two muon decay channels seems close to the sensitivities of future  BESIII searches where $10^{10}$ hyperon pairs produced in charmonium decays are expected \cite{Li:2016tlt}. Notice however that the  loop-model is not valid for heavy neutrino mass scales. The maximum rates for the $e\mu$ and $\mu\mu$ channels consistent with this scenario are far away from any current or future possible detection as it is shown in Table \ref{BRs}. It is worth mentioning that according to \cite{Fuks:2020zbm}, searches for high $p_T$ signals of LNV at the LHC may eventually be sensitive to effective muon Majorana masses of the order of a few GeV.
  
In addition, we also considered the study of short-range contributions in two appealing scenarios. First, we consider a specific low-scale seesaw model which includes two Majorana neutrinos with masses in the TeV range and not very suppressed mixings \cite{Ilakovac:1994kj, Hernandez-Tome:2019lkb}, and three light active neutrinos. In this model, the lepton number violating effects are encoded in the mass splitting of the heavy neutrinos ($r\not = 1$), while the heavy-light mixing angles are bounded from the perturbative unitarity condition. 
Second, we consider the contribution of the HTM, which can generate neutrino masses through the type-II seesaw mechanism, and contains a doubly charged scalar that couples to equal-sign leptons. Using current bounds on the parameters of both models,  we find that the branching fractions for $\Sigma^-\to pee$ are less suppressed than in the loop-mechamism discussed above, but still far below the current and expected sensitivities at BESIII.
We conclude that, even if the sensitivities of BESIII are pushed to their extreme expectations, results for the two electron channels of neutrinoless doble beta hyperon decays will  not be competitive with nuclear neutrinoless doble beta decays. 

\section*{Acknowledgements}

GHT would like to thank PROGRAMA DE BECAS POSDOCTORALES DGAPA-UNAM and DGAPA-PAPIIT UNAM, grant no. IN110622 for financial support. GHT also acknowledges Prof. Pablo Roig for partial support at the beginning of this work through catedra Marcos Moshinsky (Fundaci\'on Marcos Moshinsky).
The work of G.L.C. was supported by Ciencia de Frontera Conacyt project No. 428218 and perfil PRODEP IDPTC 162336. DPS acknowledges CONACYT for the financial support received during two years to obtain his Master's degree. We also wish to thank E. Peinado for helpful discussion.   

\appendix

\section*{One-loop functions}

The one loop functions are verified with package \textit{Package-X} \cite{Patel:2015tea}. For completeness, we report the definition and decomposition of the two and four-point functions appearing in the computations described in the main text. 
\begin{widetext}
\eq{
\frac{i}{16\pi^2}\left\{B_0, B_\mu \right\}(p_1^2,m_0^2,m_1^2)=&\mu^{4-\textrm{D}}\int \frac{d^Dq}{(2\pi)^D} \frac{\left\{1,\, q_\mu\right\}}{[q^2-m_0^2][(q+p_1)^2-m_1^2]},\\
\frac{i}{16\pi^2}\left\{D_0, D_\mu \right\}\left(\textrm{args}\right)=
&\mu^{4-\textrm{D}}\int \frac{d^Dq}{(2\pi)^D}\frac{\left\{1,\, q_\mu\right\}}{[q^2-m_0^2][(q+p_1)^2-m_1^2][(q+p_2)^2-m_2^2][(q+p_3)^2-m_3^2]},
}
\end{widetext}
where $D=4-2\epsilon$, and we have defined $(\textrm{args})=(p_{10},p_{12},p_{23},p_{30},p_{20},p_{13},m_0^2,m_1^2,m_2^2,m_3^2)$. The decomposition of the vectorial functions is given by
\eq{
B_\mu(p_1^2,m_0^2,m_1^2)=&p_{1\mu}B_1(p_1^2,m_0^2,m_1^2),\\ D_\mu\left(\textrm{args}\right)=& \sum_{i=1}^3p_{i\mu}D_i\left(\textrm{args}\right),
}with $p_{ij}=(p_i-p_j)^2$, and $p_0=0$.
Analytical expressions for the two point functions can be derived directly with \textit{Package-X}. Then, the relevant factors appearing in eqs. (\ref{v-two-points-functions}) and (\ref{a-two-points-functions}) are given by
\begin{widetext}
\eq{
C_{v0}^{\eta j}=\frac{i\kappa_{v_-}(0)}{16\pi^2}\Bigg[\frac{1}{2t}\left(m_{\nu_j}^2-m_\eta^2+t\right)\log \left(\frac{m_\eta^2}{m_{\nu_j}^2}\right)+\Lambda(t,m_\eta^2,m_{\nu_j}^2)+2-\log \left(\frac{m_\eta^2}{\mu^2}\right)+\Delta\Bigg],
}
\eq{
C_{vA}^{\eta j}&=-C_{v1}^{\eta j}=\frac{i\kappa_{v_+}(0)}{16\pi^2}\Bigg[\frac{1}{4t^2}\left(2 m_{\eta}^2m_{\nu_j}^2+2m_{\nu_j}^2 t+t^2-m_{\eta}^4-m_{\nu_j}^4\right)\log \left(\frac{m_\eta^2}{m_{\nu_j}^2}\right)\nonumber\\&+\frac{1}{2t}\left(m_{\eta}^2-m_{\nu_j}^2\right)\left(1+\Lambda(t,m_\eta^2,m_{\nu_j}^2)\right)+\frac{1}{2}\Lambda(t,m_\eta^2,m_{\nu_j}^2)+1-\frac{1}{2}\log \left(\frac{m_\eta^2}{\mu^2}\right)+\frac{\Delta}{2}\Bigg],
}
\end{widetext}
with $\Delta=\frac{1}{\epsilon}-\gamma_E+\log\left(4\pi\right)$, $\Lambda(a,b,c)=\frac{\lambda(a,b,c)}{2a}\int_0^1 dx [ax^2+(b+c-a)x+b]^{-1}$, and $\lambda$ is the so-called K\"allen function. In the limit $m_{\nu_j}\to 0$, useful for the results in the \textit{scenario A}, the above expressions reduce to
\begin{widetext}
\eq{
C_{v0}^{\eta 0}=\frac{i\kappa_{v_-}(0)}{16\pi^2}\Bigg[2-\frac{\left(t-m_{\eta}^2\right)}{t}\log\left(1-\frac{t}{m_{\eta}^2}\right)-\log\left(\frac{m_\eta^2}{\mu^2}\right)+\Delta\Bigg],
}
\eq{
C_{vA}^{\eta 0}&=-C_{v1}^{\eta 0}=\frac{i\kappa_{v_+}(0)}{16\pi^2}\Bigg[1+\frac{m_{\eta}^2}{t}-\frac{\left(t^2-m_{\eta}^4\right)}{2t^2}\log\left(1-\frac{t}{m_{\eta}^2}\right)-\frac{1}{2}\log \left(\frac{m_\eta^2}{\mu^2}\right)+\frac{\Delta}{2}\Bigg].
}
\end{widetext}

Cumbersome analytical expressions in terms of Log and Dilog functions are obtained for the factors in eqs. (\ref{v-four-points-functions}) and (\ref{a-four-points-functions}) in the type-pole approximation. We have evaluated them numerically with the help of the package\textit{ Collier} \cite{Denner:2016kdg}.

Finally, the contribution of the diagram obtained from the exchange of final charged leptons in  (\ref{amplitude-2}) is given by replacing the leptonic and hadronic currents with the following:
\begin{widetext}
\eq{
L^{\alpha\beta}_2&\equiv\bar{u}(p_1)\gamma^{\alpha}(1-\gamma_5)\gamma^{\beta}v(p_2),\\
H_{2\alpha\beta}&=\sum_{\eta,j} m_{\nu_j}U_{\ell_1 j}U_{\ell_2 j}\bigg\{
\bar{u}(p_B)\gamma_{\alpha}\left[\left(C_{2v_0}^{\eta j}+ C_{2a_0}^{\eta j}\gamma_5 \right)m_{\eta}+ \left(C_{2v_1}^{\eta j}+ C_{2a_1}^{\eta j}\gamma_5 \right)\cancel{p}_1\right.\nonumber\\ & +\left. \left(C_{2v_2}^{\eta j }+ C_{2a_2}^{\eta j}\gamma_5 \right)\cancel{p}_2+\left(C_{2v_A}^{\eta j}+ C_{2a_A}^{\eta j}\gamma_5 \right)\cancel{p}_A \right]\gamma_{\beta}u(p_A)\bigg\}\ .
} 
\end{widetext}
The $C_{2vr}^{\eta j}$ and $C_{2ar}^{\eta j}$ functions can be obtained from those reported in section \ref{hff} with the replacements: 
\eq{
C_{2v0}=&\,C_{1v0}\, (t\leftrightarrow u),\quad C_{2a0}=C_{1a0}   (t\leftrightarrow u),\nonumber\\
C_{2vA}=&\,C_{1vA}\, (t\leftrightarrow u),\,\,\,\,C_{2aA}=C_{1aA} (t\leftrightarrow u),\nonumber\\
C_{2v1}=&\,C_{1v2}\, (t\leftrightarrow u),\quad C_{2a1}=C_{1a2}   (t\leftrightarrow u),\nonumber\\
C_{2v2}=&\,C_{1v1}\, (t\leftrightarrow u),\quad C_{2a2}=C_{1a1}   (t\leftrightarrow u).
}

\section*{Kinematics}

The differential partial width for the short-range contributions is given by 
\eq{
d\Gamma=\frac{1}{(2\pi)^3}\frac{\vert \bar{\mathcal{M}}\vert^2}{32m_{A}^3\cdot z}du\,dt,
}where $z=1$\,(2) for  $\ell_1\neq \ell_2$ ($\ell_1=\ell_2$) channels, and the integration limits
\eq{
t_{\{min, max\}}&=m_A^2+m_B^2-u+\frac{1}{2u}\bigg[(m_A^2-m_2-u)\\
&\,(u-m_B^2+m_1^2)\mp\nonumber\\&
\,[\lambda(m_A^2,m_2^2,u) \lambda(u,m_B^2,m_1^2)]^{1/2}\bigg],\nonumber
}with $\lambda$ the triangle function, and
\eq{
u_{\min}=(m_B+m_1)^2,\quad u_{\max}=(m_A-m_2)^2.
}
The amplitude squared can be written as follows 
\eq{
\vert\mathcal{M}\vert^2=G_F^4 \frac{[C_{3}^{XYZ}]_{\ell_1\ell_2}^2}{\Lambda^2}F_K,
}the above expression is valid for both $C_{3}^{LLR}$  and $C_{3}^{LLL}$ Wilson coefficients in eqs. (\ref{WCHN}) and (\ref{WCHT}), respectively. 
\eq{
F_K&=-2(m_A^2+m_B^2-t-u)[A^2(m_1^2+m_2^2-2m_A m_B\nonumber\\
&-t-u)+B^2(m_1^2+m_2^2+2m_A m_B-t-u)]. 
}


\end{document}